\documentclass[12pt,a4paper]{article}
\usepackage{mathrsfs}
\usepackage{epsfig}
\begin{document}
\title{ The anomalous top quark coupling $tqg$ and $tW$ production at the $LHC$}
\author{Chong-Xing Yue, Jue Wang, You Yu, Ting-Ting Zhang\\
%\thanks{E-mail:cxyue@lnnu.edu.cn} \\
{\small Department of Physics, Liaoning  Normal University, Dalian
116029, P. R. China}
\thanks{E-mail:cxyue@lnnu.edu.cn}}
\date{\today}

\maketitle
\begin{abstract}
Many new physics models beyond the standard model ($SM$) can give rise to the large
anomalous top couplings $tqg$ ($q=u$ and $c$). We focus our attention on these couplings
induced by the topcolor-assisted technicolor ($TC2$) model and the littlest Higgs model
with $T$-parity (called $LHT $ model), and consider their contributions  to the
production cross section and the charge asymmetry for $tW$ production at the $LHC$.
We find that the anomalous top coupling $tqg$ induced by these two kinds of new physics models can indeed generate sizable
charge asymmetry. The correction effects of the $LHT $ model on the production cross sections of the processes
$pp\rightarrow tW^-+X$ and $pp\rightarrow \bar{t}W^++X$ are significant large, which might be detected at the $LHC$.
\vspace{1cm}

 \vspace{2.0cm} \noindent
 {\bf PACS numbers}: 14.65.Ha, 12.60.Cn, 12.15.Lk

\end{abstract}
\newpage
\noindent{\bf 1. Introduction }\vspace{0.5cm}

One of the main goals of the current or future high energy
experiments, such as the $LHC$ and $ILC$,  is to search for new
physics beyond the standard model ($SM$) [1]. Because of the largest
mass of the top quark among all observed particles within the $SM$,
it may be more sensitive to new physics than other fermions and it
may serve as a window to probe new physics. Thus, studying the
correction effects of new physics on observables about top quark is
a good way to test the $SM$ flavor structure and to learn more about
the nature of electroweak symmetry breaking ($EWSB$) [2].

In the $SM$, top quark can be produced singly via electroweak
interaction at hadron colliders. At leading order, there are three
kinds of the partonic processes: the s-channel process
($q'\bar{q}\rightarrow tb$) involving the exchange of a time-like $W$
boson, the t-channel process ($bq\rightarrow tq'$) involving the
exchange of a space-like $W$ boson, and the $tW$ production process
 $(gb\rightarrow tW^-)$ involving an on-shell W boson. These
processes have completely different kinematics and can be observed
separately [2]. Furthermore, the t-channel process is the main source
of single top production, both at the $Tevatron$ and the $LHC$. At
the $Tevatron$, the contributions of the $tW$ production process are
very small, while the contributions from the s-channel production
process are very small at the $LHC$. Thus, an accurate description
of all the three production processes is important.

$tW$ production at hadron colliders has been calculated at next leading order ($NLO$) in the $SM$ [3]
and been extensively studied in Refs.[4, 5].
It has been shown that this process is
observable at the $LHC$ using the fully simulated data at the $CMS$
and $ATLAS$ detectors [6, 7]. In the $SM$, the $tW$ production
channel is charge symmetric, which means that the production cross
section for the process $pp\rightarrow tW^-+X$ is equal to that for
the process $pp\rightarrow \bar{t}W^++X$. However, the charge
asymmetry in the $tW$ production process can be generated by
non-$SM$ values of $V_{td}$ and $V_{ts}$ of $CKM$ matrix [8] and  by
the anomalous top coupling $tqg$ ($q=u$ or $c$) [9].

In the $SM$, the anomalous top quark coupling $tqg$ is
absent at tree level and is extremely  suppressed at one loop due to
the $GIM$ mechanism [10], which can not be detected in current or
future high-energy experiments. However, it may be large in some new physics models
 beyond the $SM$, such as the topcolor-assisted technicolor ($TC2$) model [11, 12],
 the littlest Higgs model with $T$-parity (called $LHT $ model) [13], etc. In this paper,
 we will focus our attention on the anomalous
top couplings induced by the $TC2$ model and the $LHT$ model, and
calculate their contributions  to the production cross section and
the charge asymmetry for $tW$ production at the $LHC$ with the center-of-mass ($c. m.$) energy $\sqrt{s}=14TeV$.
Our numerical results show that the contributions of the anomalous
top coupling $tqg$ induced by the $TC2$ model to the $tW$ process
are generally smaller than those for the $LHT$ model. With
reasonable values of the free parameters of the $LHT$ model, its
corrections to the production cross sections of the processes
$pp\rightarrow tW^-+X$ and $pp\rightarrow \bar{t}W^++X$ are in the ranges of
$14\% \sim 32\%$ and $11\% \sim 24\%$, respectively. The value of the charge asymmetry
parameter $R=\sigma(tW^-)/\sigma(\bar{t}W^+)$ can reach $1.05$.

After discussing the anomalous top couplings $tqg$ induced by the
$TC2$ model and the $LHT$ model, we calculate the additional
contributions of these anomalous top couplings to the $tW$
production channel at the $LHC$ in
sections 2 and 3. Our conclusions  are given in
section 4.

\vspace{0.5cm} \noindent{\bf 2. The $TC2$ model and $tW$ production
at the $LHC$ }

\vspace{0.5cm} The $TC2$ model [11] is one of the phenomenologically
viable models, which has almost all essential features of the
topcolor scenario [12]. This model has two separate strongly
interacting sectors in order to explain $EWSB$ and the large top
mass. Technicolor interaction is responsible for most of $EWSB$
via the condensation of technifermions, but contributes very little
to the top mass $\varepsilon m_t$ with the parameter
$\varepsilon\ll1$. The topcolor interaction generates the bulk of
$m_t$ through condensation of top pairs $<t\bar{t}>$, but makes only
a small contribution to $EWSB$.

The $TC2$ model predicts the existence of a number of new scalar
states at the electroweak scale: three top-pions ($\pi^\pm_t,
\pi^0_t$), a top-Higgs ($h^0_t$),  and a techin-Higgs
($h^0_{tc}$), which are bound-states of the top quark, the bottom
quark and of the techin-fermions. Since the topcolor interaction
is not flavor-universal and mainly couples to the third generation
fermions, the couplings of top-pions or top-Higgs  to the three
family fermions are non-universal, and they have large $Yukawa$
 couplings to the third generation and can induce flavor changing
($FC$) couplings. The couplings of the top-pions
($\pi^0_t,\pi^\pm_t$) to ordinary fermions, which are related to our
calculation, can be written as [11, 12, 14]
\begin{eqnarray}
\frac{m_t}{\sqrt{2}F_t}
\frac{\sqrt{\nu^2_W-F_t^2}}{\nu_W}(iK_{UL}^{tt^*}
K^{tt}_{UR}\bar{t}_Lt_R\pi^0_t+\sqrt{2}K_{UR}^{tt^*}K^{bb}_{DL}\bar{t}_Rb_L\pi^+_t+
\\\nonumber\\
iK_{UL}^{tt^*}
K^{tc}_{UR}\bar{t}_Lc_R\pi^0_t+\sqrt{2}K_{UR}^{tc^*}K^{bb}_{DL}
\bar{c}_Rb_L\pi^+_t+h.c.),
\end{eqnarray}
where  $\nu_W=\nu/\sqrt{2}\approx 174GeV$, $F_t\approx 50GeV$ is the
physical top-pion decay constant, which can be estimated from the
Pagels-Stokar formula. To yield a realistic form of the $CKM$ matrix
$V_{CKM}$, it has been shown that the values of the matrix
 elements $K^{ij}_{UL(R)}$ can be taken as [14]
\begin{eqnarray}
K_{UL}^{tt}\approx K_{DL}^{bb}\approx 1 ,
K_{UR}^{tt}\approx1-\varepsilon , K_{UR}^{tc}\leq\sqrt{2\varepsilon
-\varepsilon ^2}.
\end{eqnarray}
In the following numerical estimation, we will assume
$K^{tc}_{UR}=\sqrt{2\varepsilon-\varepsilon^2}$ and take
$\varepsilon$ as free parameter.

\begin{figure}[htb]
\vspace{0.5cm}
\begin{center}
 \epsfig{file=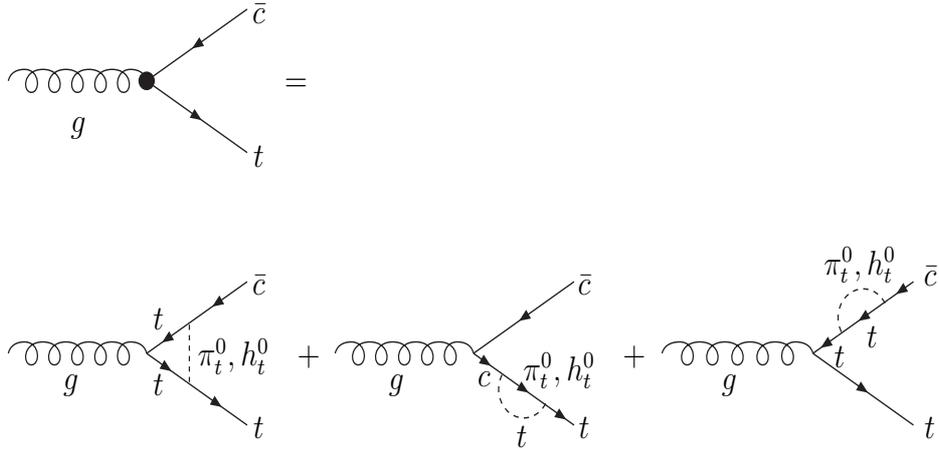,width=350pt,height=170pt}
\vspace{0.5cm} \caption{ Feynman diagrams for the effective vertex
$t\bar{c}g$ in the TC2 model.} \label{ee}
\end{center}
\end{figure}

The relevant couplings for the top-Higgs  $h^0_t$ are similar
with those of the neutral top-pion $\pi^0_t$ [14]. However, the
coupling $h^0_{tc}t\bar{t}$ is very small, which is proportionate to
a factor of $\varepsilon/\sqrt{2}$ [15].  Furthermore, the mass of
the techni-Higgs $h_{tc}$ is at the order of $1TeV$. Thus, the
contributions of $h_{tc}$ to the $tW$ production process can be
safely neglected.

From the above discussions we can see that the neutral top-pion
$\pi^0_t$ and the top-Higgs $h^0_t$ can generate the anomalous
top coupling vertex $t\bar{c}g$, which are shown in $Fig.1$. It is obvious
that the effective vertex $tcg$ can generate additional
contributions to the $tW$ production channel at the $LHC$. The
relevant Feynman diagrams are shown in $Fig.2$.

Certainly, the neutral scalars $\pi^0_t$ and $h^0_t$ can also
generate the anomalous top coupling vertex $t\bar{u}g$ via the $FC$ couplings
$\pi_t^0(h_t^0)t\bar{u}$ . However, it has been argued that the
maximum $FC$ mixing occurs between the third and second generation
fermions, and the $FC$ couplings $\pi^0_t(h^0_t)t\bar{u}$ is very
small which can be neglected [14]. Similar to $\pi^0_t$, the charged
top-pions $\pi^\pm_t$ can also give rise to the anomalous top coupling
$tcg$ via the $FC$ couplings $\pi^\pm_tbc$. However, compared with
those of $\pi^0_t$, the contributions of $\pi^\pm_t$ to the $tcg$
coupling are approximately suppressed by the factor $m^2_b/m^2_t$,
which can be safely neglected. Hence, in the following numerical
estimation, we will ignore the contributions of $\pi^\pm_t$ to the
$tW$ production process.

\begin{figure}[htb]
\vspace{0.0cm}
\begin{center}
 \epsfig{file=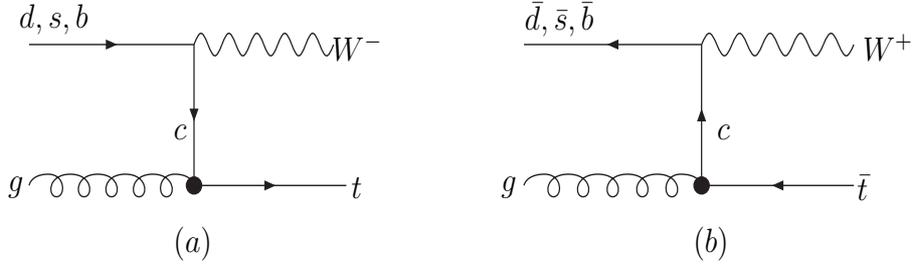,width=340pt,height=100pt}
\vspace{0.5cm} \caption{Feynman diagrams for the $tW$ production
process at the $LHC$ contributed by \hspace*{1.8cm}the anomalous top coupling
$tcg$.} \label{ee}
\end{center}
\end{figure}

One of the authors for this paper has discussed the anomalous top
coupling $tcg$ induced by the $TC2$ model in Ref.[16]. The
explicit expressions for the effective vertex $t\bar{c}g$ has been
given in Ref.[16]. In this paper, we will use $Loop Tools$ [17] and the $CTEQ6L$ parton distribution
functions ($PDFs$) [18]  to
calculate the contributions of the $TC2$ model to the $tW$
production process. The renormalization and
factorization scales ($\mu_R$ and $\mu_F$) have been taken equal to
$\mu_F=\mu_R=m_t+m_W$. The masses of the top quark and the gauge
boson $W$ are taken as $m_t=170.9GeV$ and $m_W=80.42GeV$ [19]. It is obvious
that the cross sections for the processes $pp\rightarrow tW^- +X$
and $pp\rightarrow \bar{t}W^+ +X$ are dependent  on the free
parameter $\varepsilon$ and the masses of the top-pion and
top-Higgs boson. From the theoretical point of view, $\varepsilon$
with value from 0.01 to 0.1 is favored [11]. In this paper we will
assume that its value is in the range of $0.03\sim0.08$. The masses
of the neutral top-pion and top-Higgs boson are model-dependent and
are usually of a few hundred $GeV$ [12]. In our numerical
estimation, we will take $m_{\pi_t^0}=m_{h_t^0}=M$ and assume that
the value of $M$ is in the range of $200GeV\sim500GeV$.

\begin{figure}[htb]
\vspace{-0.5cm}
\begin{center}
 \epsfig{file=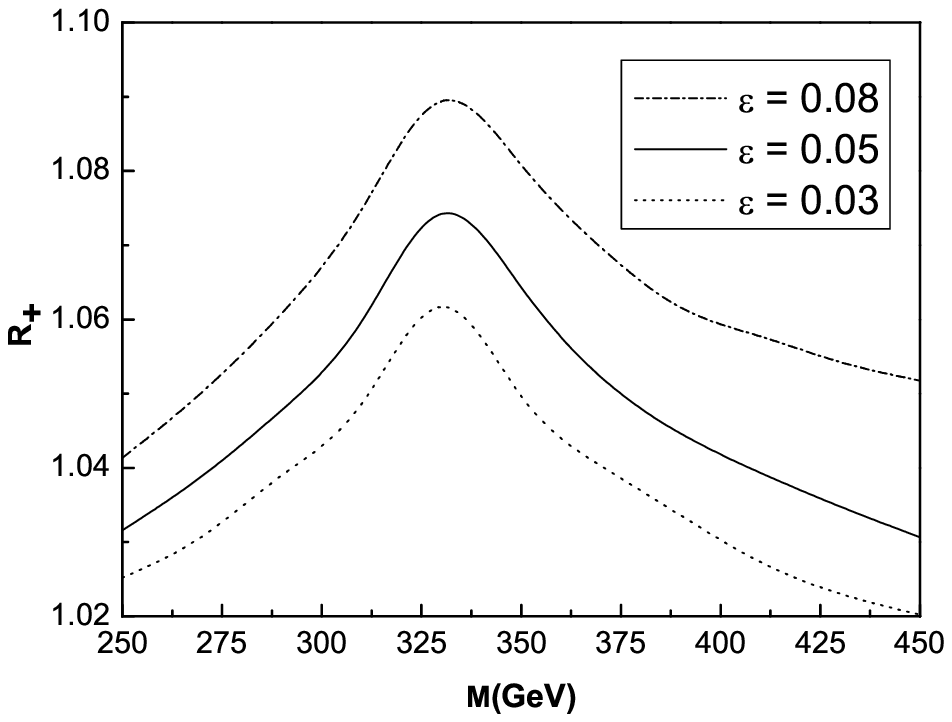,width=220pt,height=200pt}
\put(-116,-10){ (a)}\put(117,-10){ (b)} \hspace{0cm}\vspace{-0.25cm}
\epsfig{file=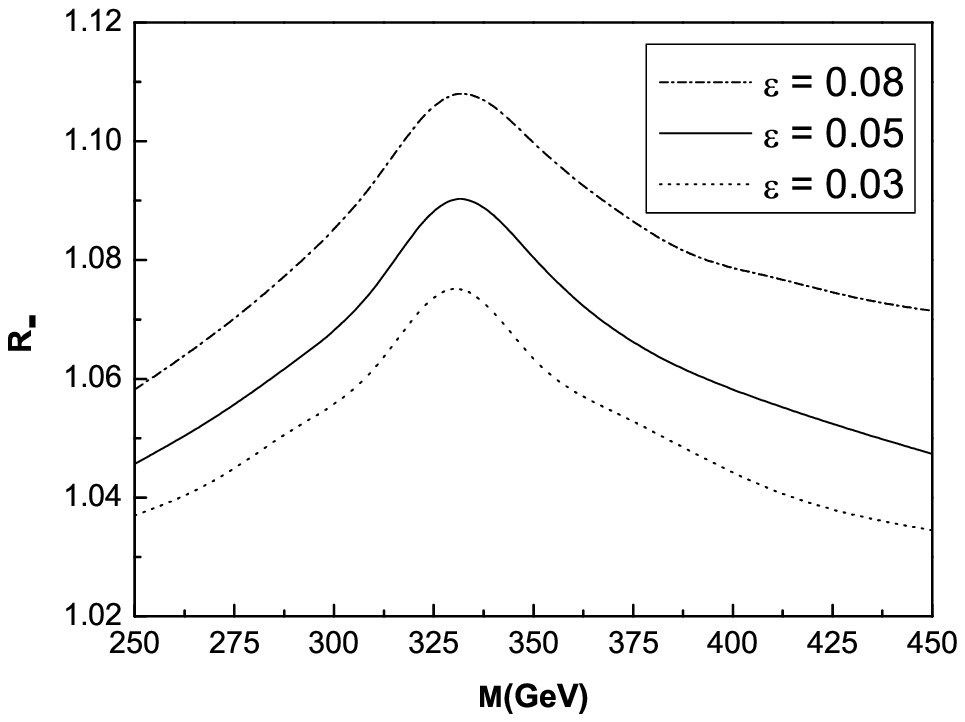,width=220pt,height=200pt} \hspace{-0.5cm}
 \hspace{10cm}\vspace{-1cm}
 \vspace{0.5cm}
\caption{The relative correction parameters $R_+(a)$ and $R_-(b)$ as
function of the mass \hspace*{2cm}parameter $M$ for three values of
the parameter $\varepsilon$.} \label{ee}
\end{center}
\end{figure}

\begin{figure}[htb]
\vspace{-0.5cm}
\begin{center}
 \epsfig{file=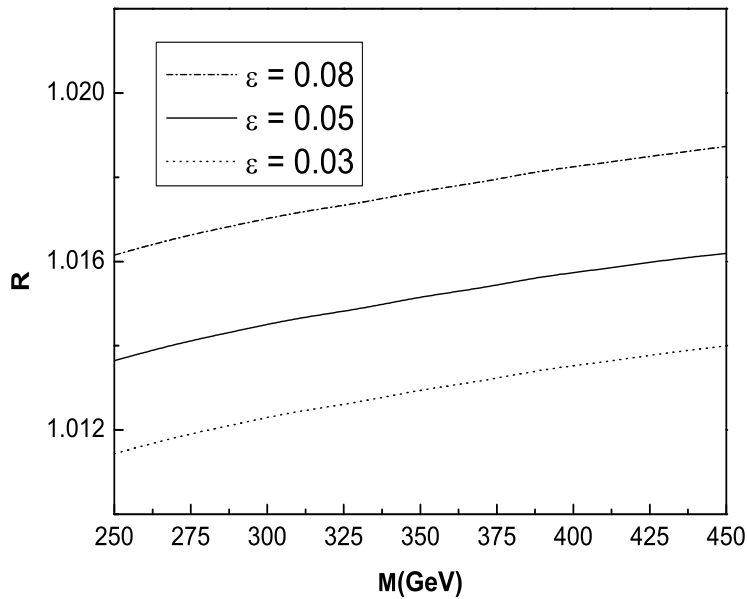,width=320pt,height=280pt}
 \vspace{-0.5cm}
 \caption{The charge asymmetry parameter $R$ as a function of $M$
 for the parameter \hspace*{2cm}$\varepsilon=0.03$, $0.05$ and $0.08$.} \label{ee}
\end{center}
\end{figure}

To see whether the contributions of the anomalous top coupling $tcg$
induced by the $TC2$ model to the $tW$ production channel  can be detected at the $LHC$, we define the relative
correction parameters as
\begin{eqnarray}
R_+=\frac{\sigma(\bar{t}W^+)}{\sigma^{SM}(\bar{t}W^+)},\hspace{0.5cm}
R_-=\frac{\sigma (tW^-)}{\sigma ^{SM}(tW^-)},
\end{eqnarray}
where $\sigma(\bar{t}W^+)$ and $\sigma(tW^-)$ denote the total
production cross sections including the contributions from the $SM$
and the $TC2$ model for the processes $pp\rightarrow \bar{t}W^+ +X$
and $pp\rightarrow tW^-+X$, respectively. The charge asymmetry
parameter $R$ is defined as $R=\sigma(tW^-)/\sigma(\bar{t}W^+)$. Since the
$PDF$ for the bottom quark in proton is same as that for the
anti-bottom quark, there is $R=1$ in the $SM$.

Our numerical results are summarized in $Fig.3$ and $Fig.4$, in
which we plot the  parameter $R_i$ as function of the mass parameter
$M$ for the $c. m.$ energy $\sqrt{s}=14TeV$ and three values of the free parameter $\varepsilon$. One can
see from $Fig.3$ that there is a peak at $M \sim 330GeV$, which is due to the effect of the $t\bar{t}$ in the loop going on-shell and  the anomalous top coupling $tcg$ increasing. In all of the parameter space of the
$TC2$ model, the value of $R_+$ is smaller than that of $R_-$ and
the value of the parameter $R$ is larger than 1, which leads to an
charge asymmetry for the $tW$ production process. For
$0.03\leq\varepsilon\leq0.08$ and $200GeV\leq M\leq500GeV$, the
corrections to the production cross sections of the processes
$pp\rightarrow\bar{t}W^++X$ and $pp\rightarrow tW^-+X$ are in the
ranges of $2.5\%\sim5.2\%$ and $3.7\%\sim7.2\%$, respectively. The
value of the charge asymmetry parameter $R$ is in the range of
$1.011\sim1.018$. It has been shown [6, 7] that the production cross
section of $tW$ production at the $LHC$ can be measured with
precision of about $9.9\%$ and $2.8\%$ for $10fb^{-1}$ and
$30fb^{-1}$ of integrated luminosity of data, respectively. Thus, it
is impossible to detect the charge asymmetry induced by the $TC2$ model for
the $tW$ production process at the $LHC$ even for the $c. m.$ energy $\sqrt{s}=14TeV$.

\vspace{0.5cm} \noindent{\bf 3. The $LHT$ model and $tW$ production
at the $LHC$ }

\vspace{0.5cm}Little Higgs theory [20] was proposed as an
alternative solution to the hierarchy problem of the $SM$, which
provides a possible kind of $EWSB$ mechanism accomplished by a
naturally light Higgs boson. In order to make the littlest Higgs
model consistent with electroweak precision tests and simultaneously
having the new particles of this model at the reach of the $LHC$, a
discrete symmetry, $T$-parity, has been introduced, which forms the
$LHT$ model. The detailed description of the $LHT$ model can be
found for instance in Refs.[13, 21, 22], and here we just want to
briefly review its essential features, which are related to our
calculation.

The $LHT$ model is based on an $SU(5)/SO(5)$ global symmetry
breaking pattern. A subgroup $[SU(2)\times U(1)]_{1}\times
[SU(2)\times U(1)]_{2}$ of the $SU(5)$ global symmetry is gauged,
and at the scale $f$ it is broken into the $SM$ electroweak symmetry
$SU(2)_{L}\times U(1)_{Y}$. $T$-parity exchanges the $[SU(2)\times
U(1)]_{1}$ and $[SU(2)\times U(1)]_{2}$ gauge symmetries. The $T
$-even combinations of the gauge fields are the $SM$ electroweak
gauge bosons $W^{a}_{\mu}$ and $A_{\mu}$. The $T$-odd combinations
are $T$-parity partners of the $SM$ electroweak gauge bosons.

After taking into account $EWSB$, at the order of $v^{2}/f^{2}$, the
masses of the $T$-odd set of the $SU(2)\times U(1)$ gauge bosons are
given  as
\begin{eqnarray}
M_{A_{H}}=\frac{g_{1}f}{\sqrt{5}}[1-\frac{5v^{2}}{f^{2}}],\hspace{0.5cm}M_{Z_{H}}\approx
M_{W_{H}}=g_{2}f[1-\frac{v^{2}}{8f^{2}}],
\end{eqnarray}
where $v=246GeV$ is the electroweak scale and $f$ is the scale
parameter of the gauge symmetry breaking of the $LHT$ model. $g_{1}$
and $g_{2}$ are the $SM$ $U(1)_{Y}$ and $SU(2)_{L}$ gauge coupling
constants, respectively.

\begin{figure}[htb]
\vspace{-0.5cm}
\begin{center}
 \epsfig{file=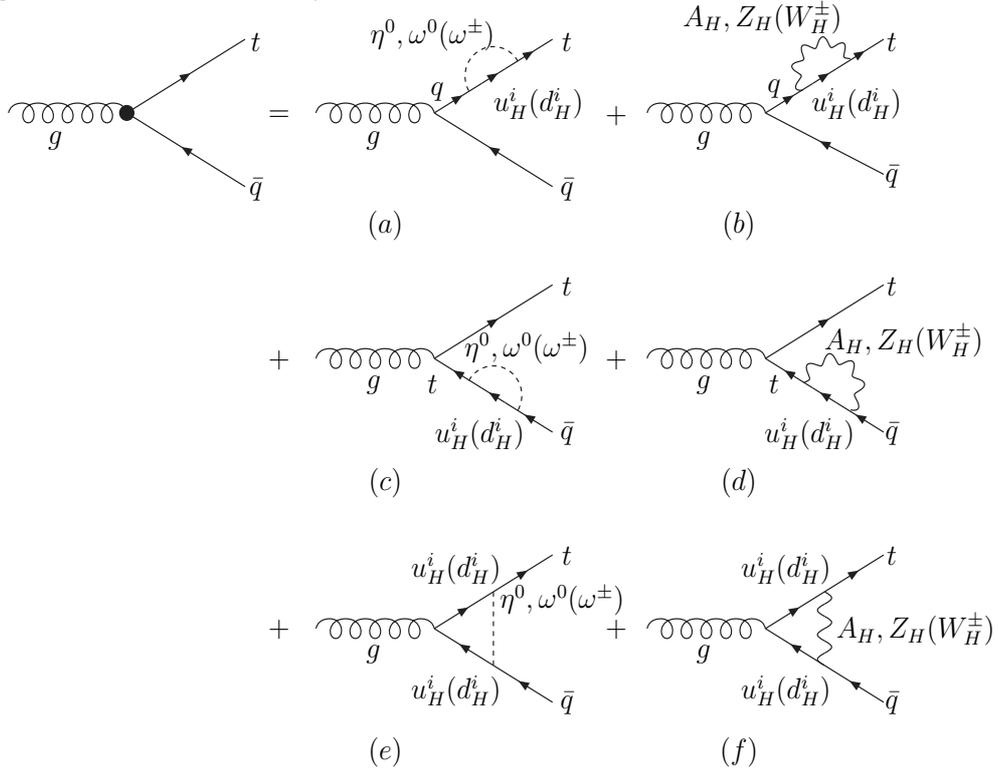,width=370pt,height=290pt}
 \vspace{-0.5cm}
 \caption{Feynman diagrams for the effective vertex $t\bar{q}g$ in the $LHT$ model.} \label{ee}
\end{center}
\end{figure}

A consistent implementation of $T$-parity also requires the
introduction of mirror fermions --- one for each quark and lepton
species. The masses of the $T$-odd (mirror) fermions can be written
in a unified manner
\begin{eqnarray}
M_{F_{i}}=\sqrt{2}k_{i}f,
\end{eqnarray}
where $k_{i}$ are the eigenvalues of the mass matrix $k$ and their
values are generally dependent on the fermion species $i$. These new
fermions ($T$-odd quarks and $T$-odd leptons) have new $FC$
interactions with the $SM$ fermions. These interactions are governed
by new
 mixing matrices $V_{Hd}$ and $V_{Hl}$ for down-type quarks and
 charged leptons, respectively. The corresponding matrices in the
 up-type quarks
 ($V_{Hu}$) and neutrino ($V_{H\nu}$) sectors are obtained by means of
 the relations
\begin{eqnarray}
V_{Hu}^{+}V_{Hd}=V_{CKM}, \hspace{0.5cm}V_{H\nu}^{+}V_{Hl}=V_{PMNS}.
\end{eqnarray}
Where the $CKM$ matrix $V_{CKM}$ is defined through flavor mixing in
the down-type quark sector, while the $PMNS$ matrix $V_{PMNS}$ is
defined through neutrino mixing.

\begin{figure}[htb]
 \vspace{0.5cm}
\begin{center}
 \epsfig{file=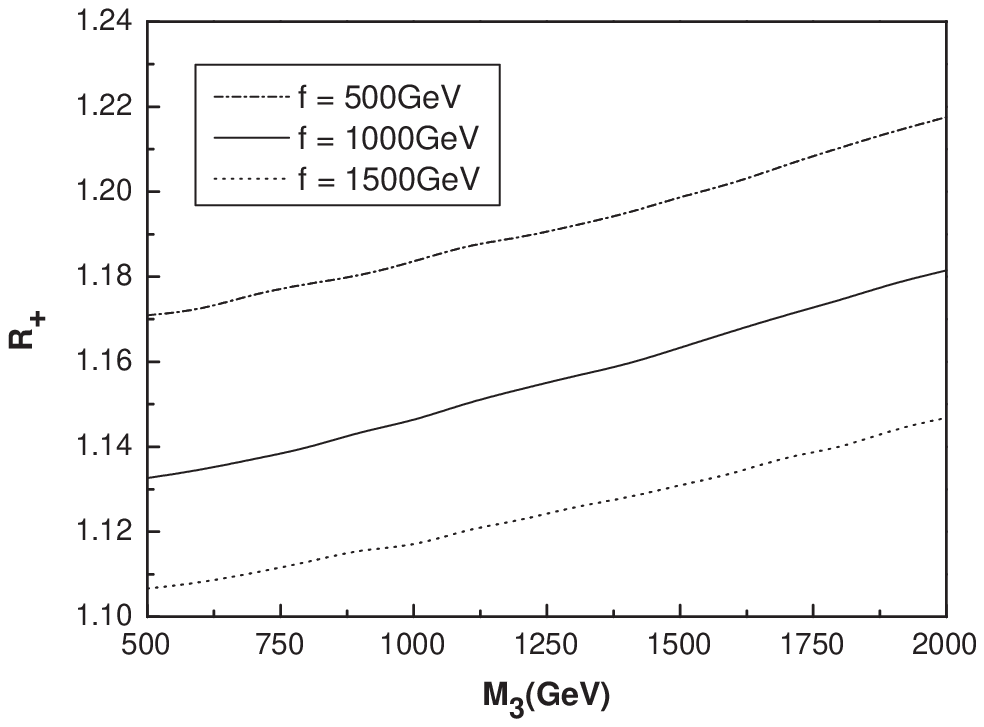,width=220pt,height=200pt}
\put(-116,-10){ (a)}\put(117,-10){ (b)} \hspace{0cm}\vspace{-0.25cm}
\epsfig{file=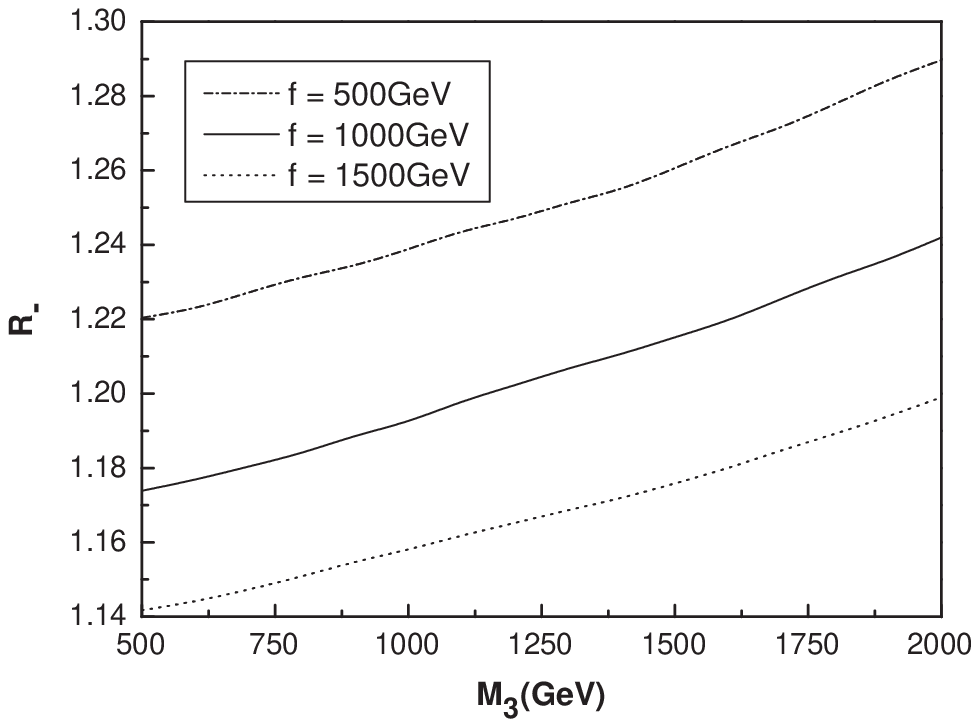,width=220pt,height=200pt} \hspace{-0.5cm}
 \hspace{10cm}\vspace{-1cm}
  \vspace{0.5cm}
\caption{In case I, the parameters $R_+(a)$ and $R_-(b)$ dependence on the mass
parameter \hspace*{1.7cm} $M_3$ for $M_1=M_2=300GeV$ and three values
of the scale parameter $f$.} \label{ee}
\end{center}
\end{figure}

The Feynman rules of the $LHT$ model have been studied in Ref.[22]
and the corrected Feynman rules of Ref.[22] are given in Refs.[23,
24]. To simplify our paper, we do not list them here.

From the above discussions, we can see that the flavor structure of
the $LHT$ model is much richer than the one of the $SM$, mainly due
to the presence of three doublets of mirror quarks and leptons and
their interactions with the ordinary quarks and leptons, which are
mediated by the $T$-odd gauge bosons ($A_H,W_H^\pm$, and $Z_H$) and
Goldstone bosons ($\eta_0$, $\omega_0$, and $\omega^\pm$). Such new
$FC$ interactions can induce the anomalous top coupling $tqg$ ($q=c$
and $u$) in quark sector. The relevant Feynman diagrams for the
effective vertex $t\bar{q}g$ are shown in $Fig.5$. To simplify our
paper, we do not give the analytical expressions of the effective
vertexes $t\bar{c}g$ and $t\bar{u}g$ here. The new coupling $tqg$
can generate significant contributions to the $FC$ top decays
$t\rightarrow cg$, $t\rightarrow cqg$ and the $FC$ single top
production processes $pp\rightarrow \bar{t}c+X$, $pp\rightarrow
t+X$, and $pp\rightarrow tg+X$ [25]. In this section, we will
consider its contributions to $tW$ production at the
$LHC$. Similar with section 2, we use  the $Loop Tools$ [17] to give
our numerical results in the $'t$ $Hooft$-Feynman gauge. In our
calculation, we use the corrected Feynman rules including the high
order $\nu^2/f^2$ terms and neglect the terms proportioning to $m_c/
m_t$ or $m_u/ m_t$.
\begin{figure}[htb]
\vspace{0.0cm}
\begin{center}
 \epsfig{file=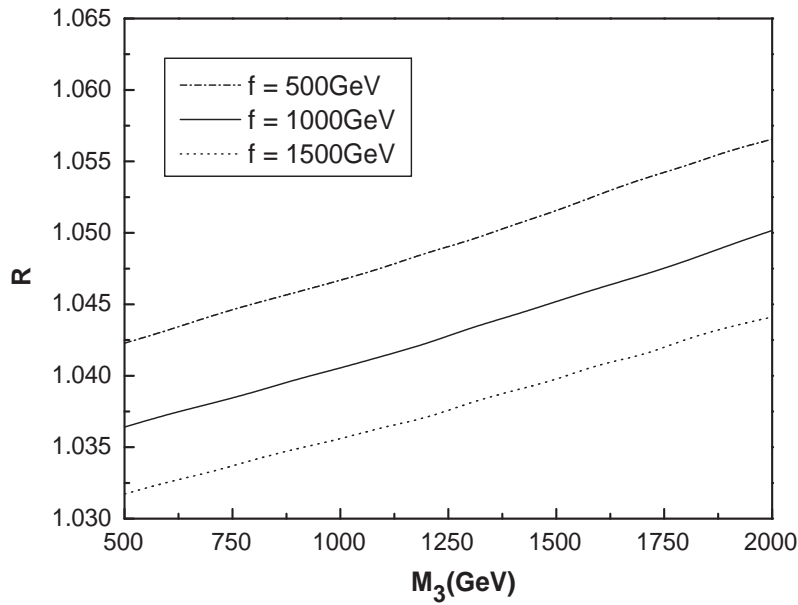,width=340pt,height=280pt}
\vspace{0.5cm} \caption{In case I, the charge asymmetry parameter $R$ as a function of the mass
parameter \hspace*{1.8cm}$M_3$ for $M_1=M_2=300GeV$ and three values
of the scale parameter $f$.} \label{ee}
\end{center}
\end{figure}

The new parameters in the $LHT$ model are the scale parameter $f$,
the mixing parameter $X_L$, the mirror fermion masses, and the
mixing matrices $V_{Hd}$ and $V_{Hl}$. The masses of the $T$-odd
gauge bosons $W^\pm_H, Z_H$, and $A_H$ can be fixed by the scale
parameter $f$. The parameter $X_L$ describes the mixing between the
$T$-even heavy top quark $T_+$ and the top quark $t$, and its value
is in the range of $0\sim1$. Since $X_L$ contributes the coupling
$tqg$ at least at order of $\nu^2/f^2$, we fix its value as 0.5. The
masses of the mirror leptons and the mixing matrix $V_{Hl}$ are not
related our calculation. For the masses of the mirror quarks, there
is $M_{U_H^i}=M_{D_H ^i}=M_i$ at $O(\nu/f)$. The mixing matrix
$V_{Hd}$ can be parameterized by three mixing angles
$\theta^d_{12},\theta^d_{23},\theta^d_{13}$ and three irreducible
phases $\delta^d_{12},\delta^d_{23},\delta^d_{13}$ [26]. The mixing
matrix $V_{Hu}$ can be determined by $V^+_{Hu}V_{Hd}=V_{CKM}$.

\begin{figure}[htb]
 \vspace{0.5cm}
\begin{center}
 \epsfig{file=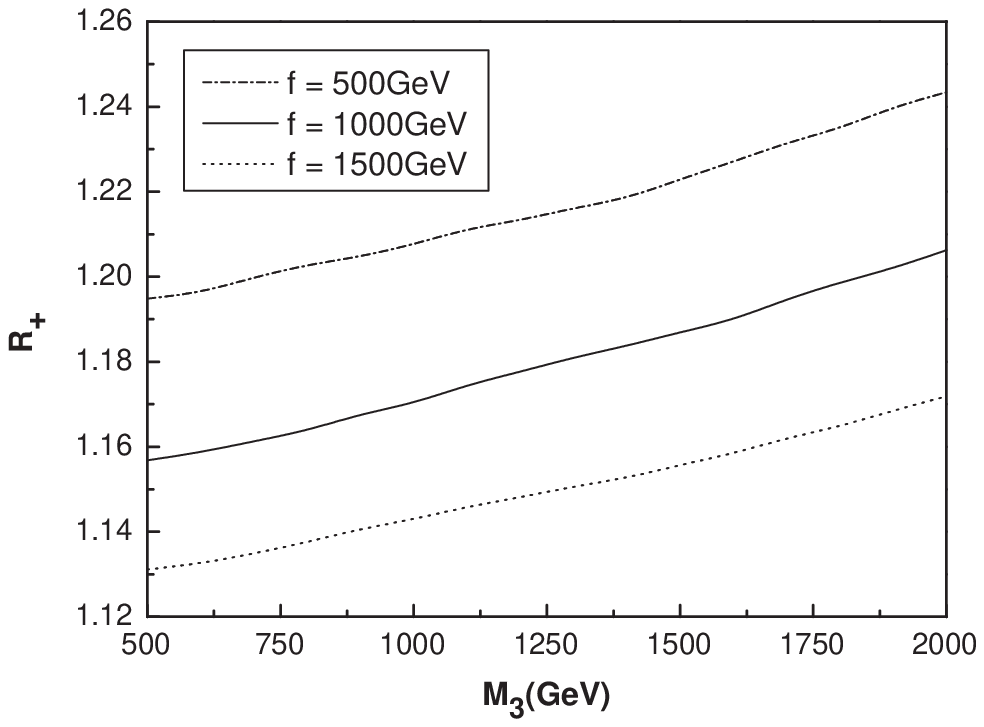,width=220pt,height=200pt}
\put(-116,-10){ (a)}\put(117,-10){ (b)} \hspace{0cm}\vspace{-0.25cm}
\epsfig{file=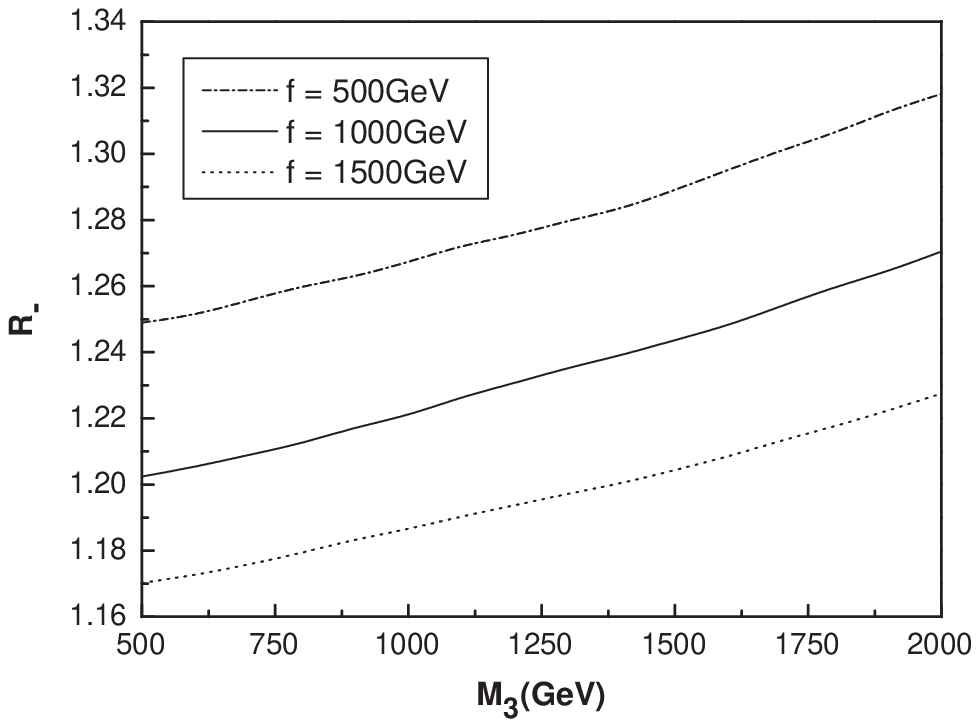,width=220pt,height=200pt} \hspace{-0.5cm}
 \hspace{10cm}\vspace{-1cm}
  \vspace{0.5cm}
\caption{Same as Fig.6 but for case II.} \label{ee}
\end{center}
\end{figure}

Refs.[21, 22, 26, 27] have studied the impact of the $LHT$ dynamics
on the $K, B$, and $D$ systems in considerable detail. They have
shown that the $LHT$ model can produce potentially sizable
effects on the relative observables and its free parameters should
be constrained. To simplify our calculation, in this paper, we only
consider two scenarios for the structure of $V_{Hd}$, which can
easily escape these constraints,

Case I: $V_{Hd}=I$, $V_{Hu}=V^+_{CKM}$,

Case II: $S^d_{23}=1/\sqrt{2} $, $S^d_{12}=S^d_{13}=0$,
$\delta^d_{12}=\delta^d_{23}=\delta^d_{13}=0.$

In both above cases, the constraints on the mass spectrum of the
mirror quarks are very relaxed. So we assume $M_1=M_2=300GeV$ and
the mass $M_3$ of the third generation mirror quarks in the range of
$500GeV\sim2000GeV$. For the scale parameter $f$, we take its
typical values, i.e. $500GeV\sim2000GeV$.

\begin{figure}[htb]
\vspace{-0.5cm}
\begin{center}
 \epsfig{file=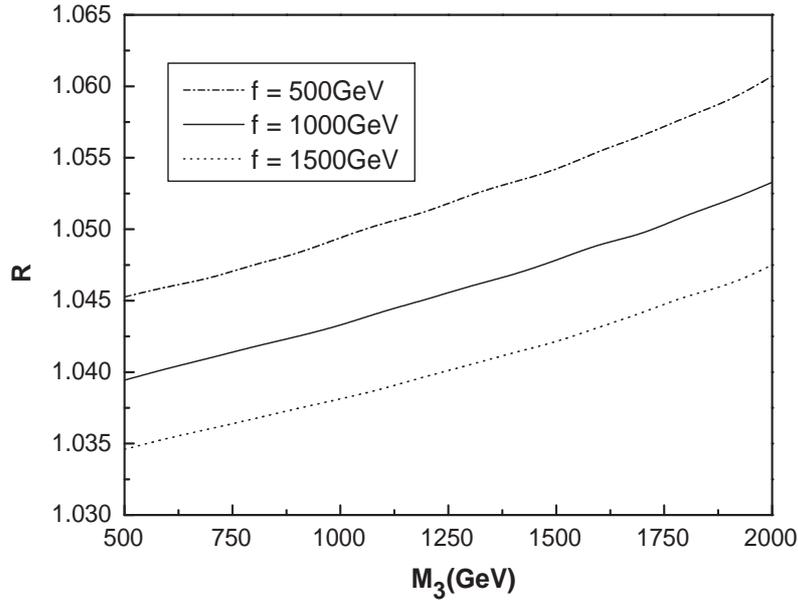,width=340pt,height=280pt}
 \vspace{-0.5cm}
 \caption{Same as Fig.7 but for case II.} \label{ee}
\end{center}
\end{figure}

The parameters $R_+, R_-$,  and $R$ contributed by the anomalous top
couplings $tcg$ and $tug$ in the $LHT$ model are plotted as
functions of the mass parameter $M_3$ for the $c. m.$ energy $\sqrt{s}=14TeV$ and three values of the scale
parameter $f$, which are shown in figures $6\sim 9$. From these figures one can see that the contributions of the anomalous top coupling
$tqg$ induced by the $LHT$ model to the $tW$ production process are generally 
larger than those for the $TC2$ model. This is partly because the
contributions of the $LHT$  model from the anomalous top couplings
$tcg$ and $tug$, while only from the anomalous top coupling $tcg$ for the $TC2$
model. The values of the parameters $R_+$, $R_-$, and the deviation $\delta R=R_- -R_+$ increase as
the mass parameter $M_3$ increases, which is because the couplings between the mirror quarks and the $SM$ quarks are proportion to the mirror quark masses. So the parameter $R$ also increases as $M_3$ increases. Certainly, compared to the parameters $R_+$ and $R_-$,  $R$ is insensitive to the mass parameter $M_3$ and its values are only in the ranges of $1.042\sim 1.056$ and $1.045\sim 1.061$ for case I and case II, respectively. These
parameters also depend on the parameterization scenarios of the
matrix $V_{Hd}$. Their values for case II are generally larger than
those for case I. In most of the parameter space of the $LHT$
model, the values of the relative correction parameters $R_+$ and
$R_-$ are larger than $1.1$. Thus, the correction effects of the
anomalous top coupling $tqg$ induced by the $LHT$ model on the $tW$
production cross section might be detected at the $LHC$. Although
the value of the charge asymmetry parameter $R$ induced by the $LHT$
model is larger than that for the $TC2$ model, its value is smaller
than $1.06$. So, observing the
charge asymmetry of $tW$ production at the $LHC$ induced
by the $LHT$ model is much challenge.

\vspace{0.5cm} \noindent{\bf 4. Conclusions }

\vspace{0.5cm}The $tW$ production process is one of important single
top production channels at the $LHC$. In the $SM$, the production
cross sections of single top quark and single anti-top quark in the
$tW$ channel are equal, i.e. $R=\sigma(tW^-)/\sigma(\bar{t}W^+)=1$.
However, the anomalous top coupling $tqg$ can generate contributions
to the cross sections $\sigma(tW^-)$ and $\sigma(\bar{t}W^+)$, and
further give rise to the charge asymmetry. If the correction effects
of the new coupling $tqg$ on the $tW$ production channel are
observed at the $LHC$, it will be helpful to test the flavor
structure of the $SM$ and further to probe new physics beyond the
$SM$.

The $TC2$ model and the $LHT$ model are two kinds of popular new
physics models, which can generate the anomalous top coupling $tqg$.
In the context of the $TC2$ and $LHT$ models, we consider the correction effects of the new
coupling $tqg$ on the $tW$ production channel at the $LHC$ with the $c. m.$ energy $\sqrt{s}=14TeV$. Our numerical results show that
they can indeed generate significant contributions to the $tW$ production process.
The contributions of the anomalous
top coupling $tqg$ induced by the $TC2$ model to the $tW$ production process
are generally smaller than those for the $LHT$ model. With
reasonable values of the free parameters for the $LHT$ model, its
corrections to the production cross sections of the processes
$pp\rightarrow tW^-+X$ and $pp\rightarrow \bar{t}W^++X$ can reach
$32\%$ and $24\%$, respectively. The value of the charge asymmetry
parameter $R=\sigma(tW^-)/\sigma(\bar{t}W^+)$ can reach $1.06$.

The $TC2$ model and the $LHT$ model can modify the $Wtb$ coupling and further produce correction effects on the $tW$ production cross section [28, 29]. However, their contributions to the production cross section of the process
$pp\rightarrow tW^-+X$ are equal to those for the production cross section of the process $pp\rightarrow \bar{t}W^++X$. Thus, such modification about the $Wtb$ coupling can not cause the charge asymmetry in the $tW$ production process at the $LHC$.

\section*{Acknowledgments} \hspace{5mm}This work was
supported in part by the National Natural Science Foundation of
China under Grants No.10975067, the Specialized Research Fund for
the Doctoral Program of Higher Education (SRFDP) (No.200801650002).

\end{document}